\documentclass[aps,pre,twocolumn,superscriptaddress,groupedaddress]{revtex4}
\usepackage{hyperref}
\usepackage{color}
\usepackage{graphics,graphicx,epsfig,wrapfig}
\usepackage{epsf,epstopdf}
\usepackage{amssymb,amsfonts,amsmath}
\usepackage{enumitem}

\usepackage{ifthen}

\newcommand\mydots{\hbox to 0.8em{.\hss.\hss.}}
\newcommand{\beqn}{\begin{eqnarray}}
\newcommand{\eeqn}{\end{eqnarray}}
\newcommand{\beq}{\begin{equation}}
\newcommand{\eeq}{\end{equation}}

\newcommand{\degree}{{}^{\rm o}}

\definecolor{ormar}{rgb}{.8,.2,0}

\newcommand{\beginsupplement}{%
        \setcounter{table}{0}
        \renewcommand{\thetable}{S\arabic{table}}%
        \setcounter{figure}{0}
        \renewcommand{\thefigure}{S\arabic{figure}}%
     }

\begin{document}

\title{A model for the integration of conflicting exogenous and endogenous signals by dendritic cells}

\author{Quentin Marcou}
\thanks{Equal contribution}
\affiliation{Laboratoire de physique th\'eorique,
    CNRS, UPMC and \'Ecole normale sup\'erieure,
    75005 Paris, France} 

\author{Irit Carmi-Levy}
\thanks{Equal contribution}
\affiliation{INSERM U932, 26 rue d'Ulm, 75005 Paris, France; Institut Curie, Section Recherche, 26 rue d'Ulm, 75005 Paris, France; Laboratoire d'Immunologie Clinique, Institut Curie, 26 rue d'Ulm, 75005 Paris, France.}

\author{Coline Trichot}
\affiliation{INSERM U932, 26 rue d'Ulm, 75005 Paris, France; Institut Curie, Section Recherche, 26 rue d'Ulm, 75005 Paris, France; Laboratoire d'Immunologie Clinique, Institut Curie, 26 rue d'Ulm, 75005 Paris, France.}

\author{Vassili Soumelis}
\thanks{Corresponding authors with equal contribution}
\affiliation{INSERM U932, 26 rue d'Ulm, 75005 Paris, France; Institut Curie, Section Recherche, 26 rue d'Ulm, 75005 Paris, France; Laboratoire d'Immunologie Clinique, Institut Curie, 26 rue d'Ulm, 75005 Paris, France.}

\author{Thierry Mora}
\thanks{Corresponding authors with equal contribution}
\affiliation{Laboratoire de physique statistique,
    CNRS, UPMC and \'Ecole normale sup\'erieure,
    75005 Paris, France}

\author{Aleksandra M. Walczak}
\thanks{Corresponding authors with equal contribution}
\affiliation{Laboratoire de physique th\'eorique,
    CNRS, UPMC and \'Ecole normale sup\'erieure,
    75005 Paris, France}

\address{}

\date{\today}

\begin{abstract}
Cells of the immune system are confronted with opposing pro- and anti-inflammatory signals. Dendritic cells (DC) integrate these cues to make informed decisions whether to initiate an immune response. Confronted with exogenous microbial stimuli, DC endogenously produce both anti- (IL-10) and pro-inflammatory (TNF$\alpha$) cues whose joint integration controls the cell's final decision. We combine experimental measurements with theoretical modeling to quantitatively describe the integration mode of these opposing signals. We propose a two step integration model that modulates the effect of the two types of signals: an initial bottleneck integrates both signals (IL-10 and TNF$\alpha$), the output of which is later modulated by the anti-inflammatory signal. We show that the anti-inflammatory IL-10 signaling is long ranged, as opposed to the short-ranged pro-inflammatory TNF$\alpha$ signaling. The model suggests that the population averaging and modulation of the pro-inflammatory response by the anti-inflammatory signal is a safety guard against excessive immune responses.
\end{abstract}

\maketitle

\section{Introduction}
Cells constantly integrate signals to adapt to their environment. In the immune system, activating signals are critical to initiate and sustain an efficient immune response, and co-exist with inhibitory signals in order to avoid excessive and uncontrolled immune responses \cite{hu2008regulation,tamayo2011pro}. Immune cells  must often integrate such opposing signals, the outcome being key to decision making between immunity versus tolerance \cite{cappuccio2015combinatorial,long2013controlling,krummel1995cd28}. This signal integration process in immune cells involves many check points that can involve kinetic proofreading \cite{McKeithan1995, Francois2013, Dushek2014} or multiple feedback loops \cite{Hoffmann2002, Tay2010}. In general, feedback allows the system to adjust its output in response to monitoring itself. Both positive and negative feedback loops have been found crucial to control the strength and duration of the system's activation in order to achieve optimal responses. Such loops represents a fundamental feature in cell development and differentiation \cite{freeman2000feedback}, hormonal homeostasis \cite{pyle2005multiple}, intracellular signalling \cite{guan2006feedback} and in the immune response \cite{hu2008regulation}. Cells can receive feedback through paracrine signals coming from their neighbours or from their own autocrine signals \cite{Youk2014, Youk2014a}. Since the adaptation to the environment occurs at the population level, autocrine and paracrine feedback may play a different role in a cell population  responding to opposing signals, notably as a function of cell density.

Dendritic cells (DC) are an essential component of the innate immune system. Acting as the body's sentinels, they are equipped with a diversity of innate receptors, including pattern recognition receptors such as Toll Like Receptors (TLRs). Engagement of TLRs by TLR ligands leads to DC maturation, a complex process which includes migration to draining lymph nodes, secretion of a diversity of chemokines and cytokines, as well as up-regulation of major histocompatibility class II (MHC-II) and co-stimulatory molecules, such as CD80 and CD86 \cite{pasare2005toll}. The latter represent crucial molecular checkpoints for orchestrating DC-T cell communication, playing a key role in the activation and expansion of CD4 T cells \cite{de2005dendritic}. 

A critical question is how the diversity of signals sensed by DC control the outcome of the DC maturation program. In this process, we can discriminate exogenous signals, i.e the nature and dose of microbial stimuli, and endogenous signals, such as autocrine factors induced by exogenous stimulation. When DC are activated by the bacterial component LPS (exogenous signal), they respond with an increased secretion of TNF-alpha (TNF$\alpha$) and interleukin (IL)-10, generally considered as prototypical pro- and anti-inflammatory signals, respectively \cite{moore2001interleukin,blanco2008dendritic}. As DC are equipped with the corresponding receptors, both TNF$\alpha$ and IL-10 act as endogenous auto-regulatory feedback signals that control the output response of the cell, and influence the final decision to initiate an immune response or not. Current and past studies have mostly studied each of these signals separately. LPS effect on DC has been extensively studied, including at various concentrations revealing dose-dependent effects \cite{verhasselt1997bacterial,dearman2009toll}. Few studies have addressed the role of the IL-10 negative feedback loop, showing that it dampens LPS-induced maturation \cite{corinti2001regulatory}. TNF$\alpha$ is a DC-activating pro-inflammatory cytokine \cite{blanco2008dendritic}, but its role as a putative positive feedback  factor on DC remains elusive. Studies of these DC-targeting regulatory signals suggest strong dependencies and cross-regulatory mechanisms between LPS, IL-10 and TNF$\alpha$, but the underlying rules remain unexplored. Mechanistic understanding requires the integrated analysis of variations in the three signals level, and their consequences on the behaviour of the system. 

In this study, we combined experimental and theoretical approaches to perform a quantitative dynamic analysis of LPS, IL-10, and TNF$\alpha$ effects on human DC. We propose an original model of the interplay between contradictory exogenous and endogenous signals in the control of DC maturation.

\section{Results}

\subsection{LPS-induced TNF$\alpha$ and IL-10 differentially control DC maturation}

\begin{figure}[h]
\noindent\includegraphics[width=\linewidth]{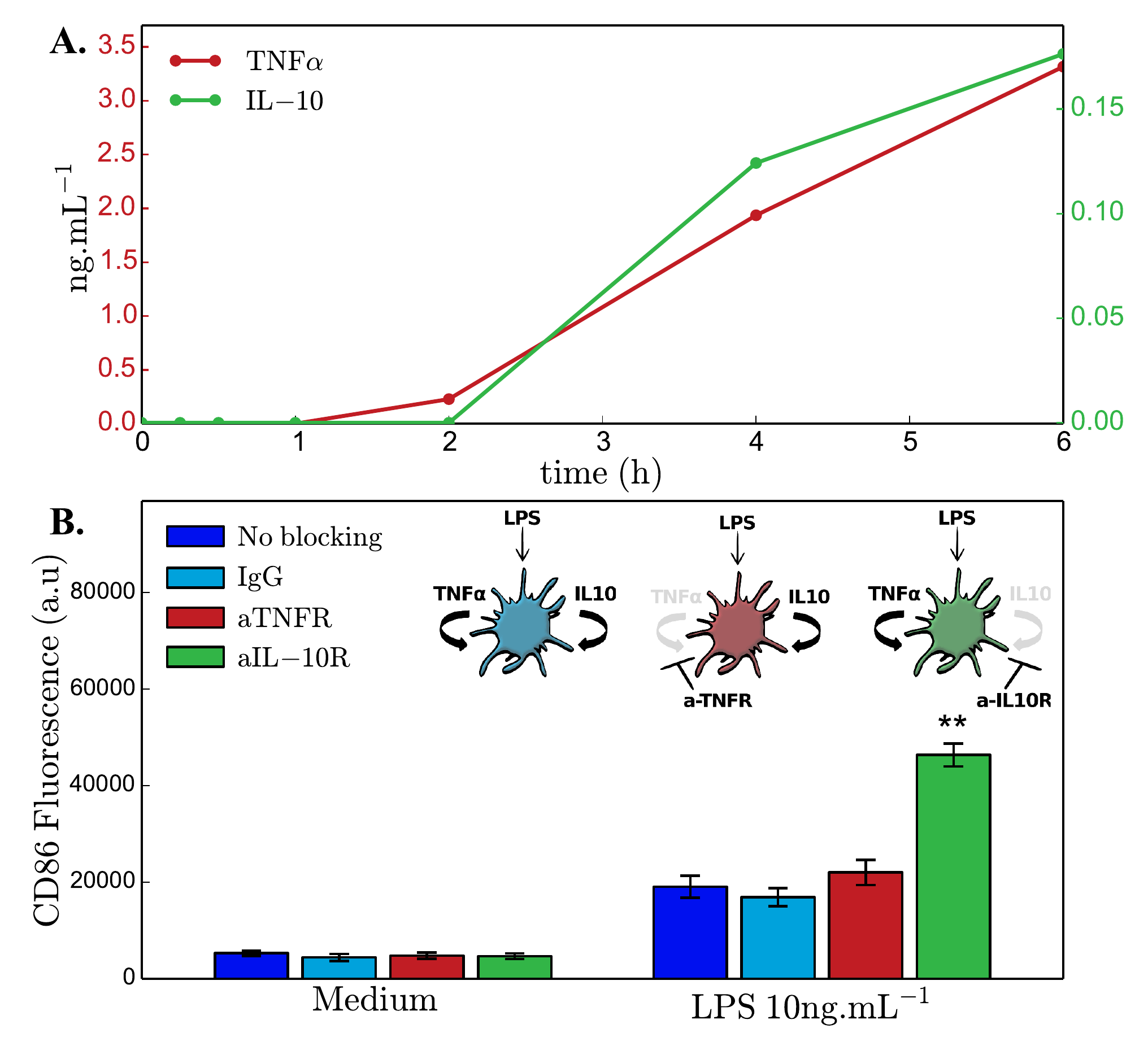}
\caption{
\textbf{LPS-induced $\mathrm{TNF \alpha}$ and IL-10 differentially control DC maturation.} \textbf{A.} Secretion of the cytokines $\mathrm{TNF \alpha}$ (red) and IL-10 (green) is monitored through time under $\mathrm{100ng.mL^{-1}}$ LPS stimulation. \textbf{B.}  CD86 fluorescence of cellular populations is increased by the presence of LPS in the medium. Blocking the regulatory loops has no effect when cells are not stimulated when blocking IL-10 pathway increases DC activation. As a control for the non blocking condition, culture with an isotypic antibody does not alter CD86 levels. Bars show the expectation of the log normal distribution, error bars the standard error of the mean of the log normal distributions. Statistical significance of the results is is assessed using Welch's t-test in logarithmic space. 
} 
\label{fig1}
\end{figure}

Upon activation by the TLR ligand LPS, DC undergo a maturation process leading to an upregulation of costimulatory molecules, such as CD86, but also production of pro- and anti-inflammatory cytokines. First, we measured the production of TNF$\alpha$ and IL-10 in response to a standard LPS concentration of 100 ng/ml. The secretion of TNF$\alpha$ was more rapid and was significant already after 2 hours, while IL-10 was detected only after 4 hours following LPS stimulation (Fig\ref{fig1}A). After 4 hours
, both cytokines were detected concomitantly in the cellular supernatant (Fig.~\ref{fig1}A). TNF$\alpha$ and IL-10 reached concentrations of $3.3$ ng/ml and $0.18$ ng/ml, respectively after 6 hours (Fig 1A).
In order to address the contribution of these two endogenous cytokines on DC maturation, we monitored CD86 using flow cytometry, in the presence and absence of blocking antibodies (Ab) to TNF$\alpha$ or IL-10(Fig.~\ref{fig1}B). LPS induced significant upregulation of CD86, consistent with an increase in DC maturation (Fig.~\ref{fig1}B).  Blocking the IL-10 loop induced a significant increase in CD86 expression. This suggested that IL-10 had a dominant negative effect in controlling LPS-induced DC maturation.

\subsection{LPS dose determines the endogenous IL-10 and TNF$\alpha$ control of DC maturation}

\begin{figure}[h]
\noindent\includegraphics[width=\linewidth]{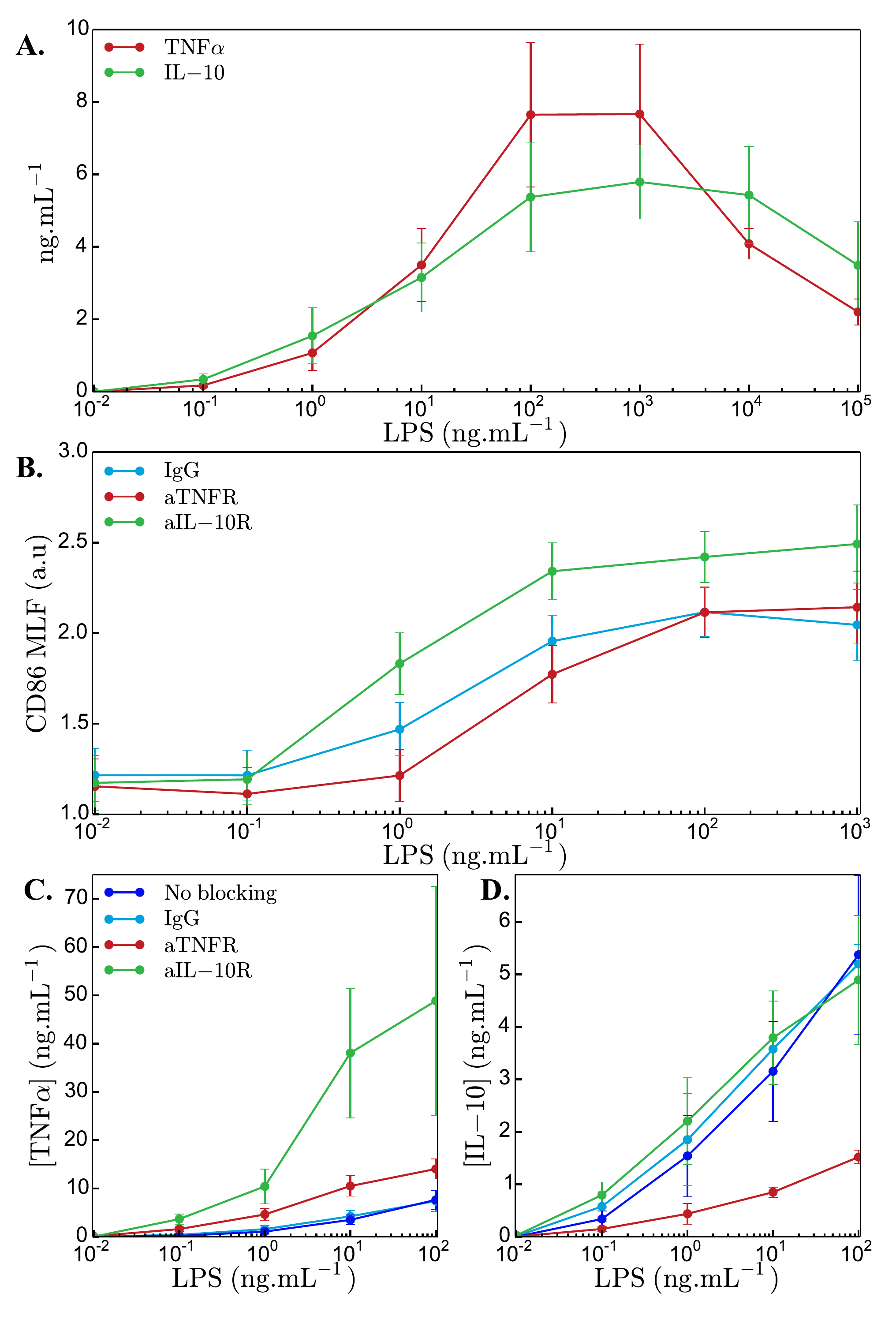}
\caption{
\textbf{LPS dose determines the endogenous IL-10 and TNF$\alpha$ control of DC maturation.}
\textbf{A.} Titration of TNF$\alpha$ and IL-10 concentrations($\mathrm{ng.mL^{-1}}$) for a wide range of LPS doses. Increasing LPS doses increase both TNF$\alpha$ (red) and IL-10 (green) secretion levels.
\textbf{B.} Activation of DC is monitored by flow-cytometry labeling the co-stimulatory molecule CD86. CD86 mean log-fluorescence (MLF) is shown for a range of LPS concentration incubated with isotypic control (blue), anti-TNFR (red) or anti-IL-10R (green) antibodies. CD86 has a sigmoidal dependence on LPS doses. Blocking IL-10 increases the maximal activation level while blocking TNF$\alpha$ decreases the sensitivity.
Cytokine response of DC in different conditions, medium (dark blue), isotypic control (blue), anti-TNFR (red), anti-IL-10R antibodies, is measured for different doses of LPS.
\textbf{C.} Blocking IL-10 increases TNF$\alpha$ secretion
\textbf{D.} Blocking TNF$\alpha$ decreases IL-10 secretion 
}
\label{fig2}
\end{figure}

Microbial-derived signals occur at various concentrations in infected tissue, in relationship to the in situ microbial load. This process is also linked to microbial clearance, which induces a local decrease in microbial signals. First, we addressed the impact of various LPS doses on endogenous TNF$\alpha$ and IL-10 production (Fig.~\ref{fig2}A). Both cytokines exhibited a similar LPS dose-dependent pattern, reaching maximum levels at a LPS concentration of 100 ng/ml (Fig.~\ref{fig2}A).
 
Given that TNF$\alpha$ and IL-10 co-exist at variable LPS concentrations, we asked whether LPS levels impact the way these endogenous signals are being integrated by DC.
To address this question, we cultured DC in the presence or absence of blocking Abs to the TNF$\alpha$ and IL-10 receptors (TNFR and IL10R) while stimulating them with different concentrations of LPS achieved by serial dilutions (Fig.~\ref{fig2}B). As for the standard LPS dose, DC maturation was quantified by CD86 expression 24 hours following LPS activation. When none of the loops were altered (no blocking or IgG control), the level of activation increased with LPS concentration and reached a plateau for sufficiently high LPS doses ($\sim100$ng/ml)  (blue curve in Fig.~\ref{fig2}B). Blocking the pro-inflammatory TNF$\alpha$ loop led to a decreased expression of CD86 (red curve in Fig.~\ref{fig2}B), while blocking the anti-inflammatory IL-10 loop led to an increased expression of CD86 (green curve in Fig.~\ref{fig2}B). However, TNF$\alpha$ loop-blocking decreased CD86 levels mostly at LPS concentrations lower than 10 ng/ml (red curve in Fig.~\ref{fig2}B). By contrast, the impact of IL-10 loop-blocking on CD86 expression was constant along a wide spectrum of medium to high LPS doses, but absent at low LPS doses (green curve in Fig.~\ref{fig2}B). The impact of the two opposite/contradictory loops differed not only in the directionality of the effect but also in mode of the effect: blocking the TNF$\alpha$ loop shifted the $EC_{50}$ (half maximal effective concentration) of the response towards higher LPS dose, while blocking the IL-10 loop affected mainly the amplitude of the response, which significantly increased in the presence of IL-10 blocking compared to its value in the absence of any blocking.

DC maturation with or without blocking the loops in the different LPS doses was quantified using the expression of a second maturation marker CD83. The expression of this marker also increased with increasing LPS dose (Fig.~\ref{figS1}). Blocking the TNF$\alpha$ loop led to a similar trend as with CD86, with a strong effect at low LPS doses, and weaker effect at high LPS doses (Fig.~\ref{figS1}). Although both maturation markers were significantly upregulated by LPS, their distribution across the DC population was different. While CD86 demonstrated a unimodal distribution, CD83 demonstrated a bi-modal one (Fig.~\ref{figS2}). In addition to surface markers, blocking the loops also had a significant effect on cytokine secretion (Fig.~\ref{fig2}C and D). Importantly, the TNF$\alpha$ and IL-10 loops reciprocally affected each other, as blocking the IL-10 loop increased TNF$\alpha$ secretion (green curve in Fig.~\ref{fig2}C compared to the other curves), and blocking of the TNF$\alpha$ loop strongly decreased IL-10 production (red curve in Fig.~\ref{fig2}D). This suggests potential cross-regulation of TNF$\alpha$ and IL-10 through DC. 	

\subsection{Modulated bottleneck model explains DC maturation control by opposing endogenous and exogenous signals}

\begin{figure}[h]
\noindent\includegraphics[height=3in]{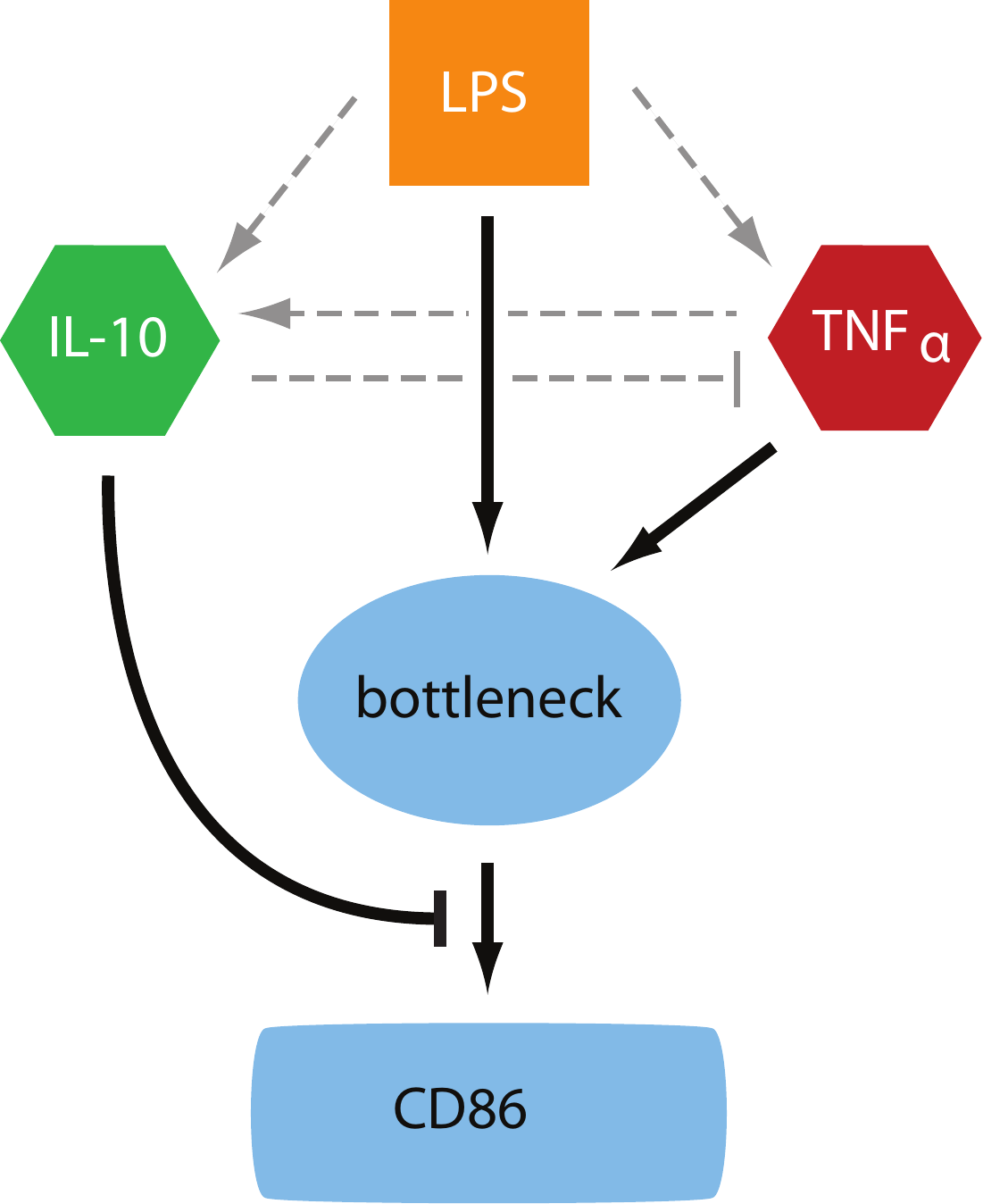}
\caption{
\textbf{Cartoon of the modulated bottleneck model.}
Arrows represent functional (not necessarily direct) interactions. LPS controls the activation of the bottleneck, as well as IL-10 and TNF$\alpha$. TNF$\alpha$ and LPS act through a common bottleneck for the activation of DC, while IL-10 modulates the activation level downstream. The model also includes partial mutual regulation of TNF$\alpha$ and IL-10. 
}
\label{fig3}
\end{figure}

In order to qualitatively understand the mechanism behind microbial-induced signal integration in DC, we used the above experimental observations to built a minimal phenomenological steady state mathematical model of CD86 response to LPS stimulation. From Fig.~\ref{fig2}B we see that the CD86 response follows a sigmoidal dependence on LPS concentration, which we denote as L and saturates at high LPS level. Additionally, both IL-10 (denoted as I) and TNF$\alpha$ (T) expressions are sigmoidal functions of LPS (Fig.~\ref{fig2}C and D and see  Materials and Methods Eqs.~\ref{eq1}-\ref{eq4}). As we noted above, TNF$\alpha$ upregulates IL-10 expression, while IL-10 downregulates TNF$\alpha$ secretion (Fig.~\ref{fig2}C and D). To avoid behavior that is not observed in the data, we assume there is a basal expression level of both TNF$\alpha$ and IL-10, even in the absence (presence) of the regulator. Results of blocking IL-10 show that additionally to repressing TNF$\alpha$, IL-10 also decreases the amplitude of the response. Lastly, it has previously been shown that TNF$\alpha$ alone, in the absence of LPS, activates and induces DC maturation. This observation suggested that TNF$\alpha$ does not just act downstream of LPS and changes the $EC_{50}$ solely by transmitting the LPS activation signal, but that TNF$\alpha$ and LPS act through a common intermediate in an additive way creating a bottleneck. This last assumption is the main idea behind our model: LPS and TNF$\alpha$ signals are integrated in the expression of one regulatory molecule. The expression of CD86 itself is not regulated directly by TNF$\alpha$ and LPS, but by the concentration and status of this central integrator (Fig.~\ref{fig3}). 

\begin{figure*}[t]
\centering
\noindent\includegraphics[width=6.5in]{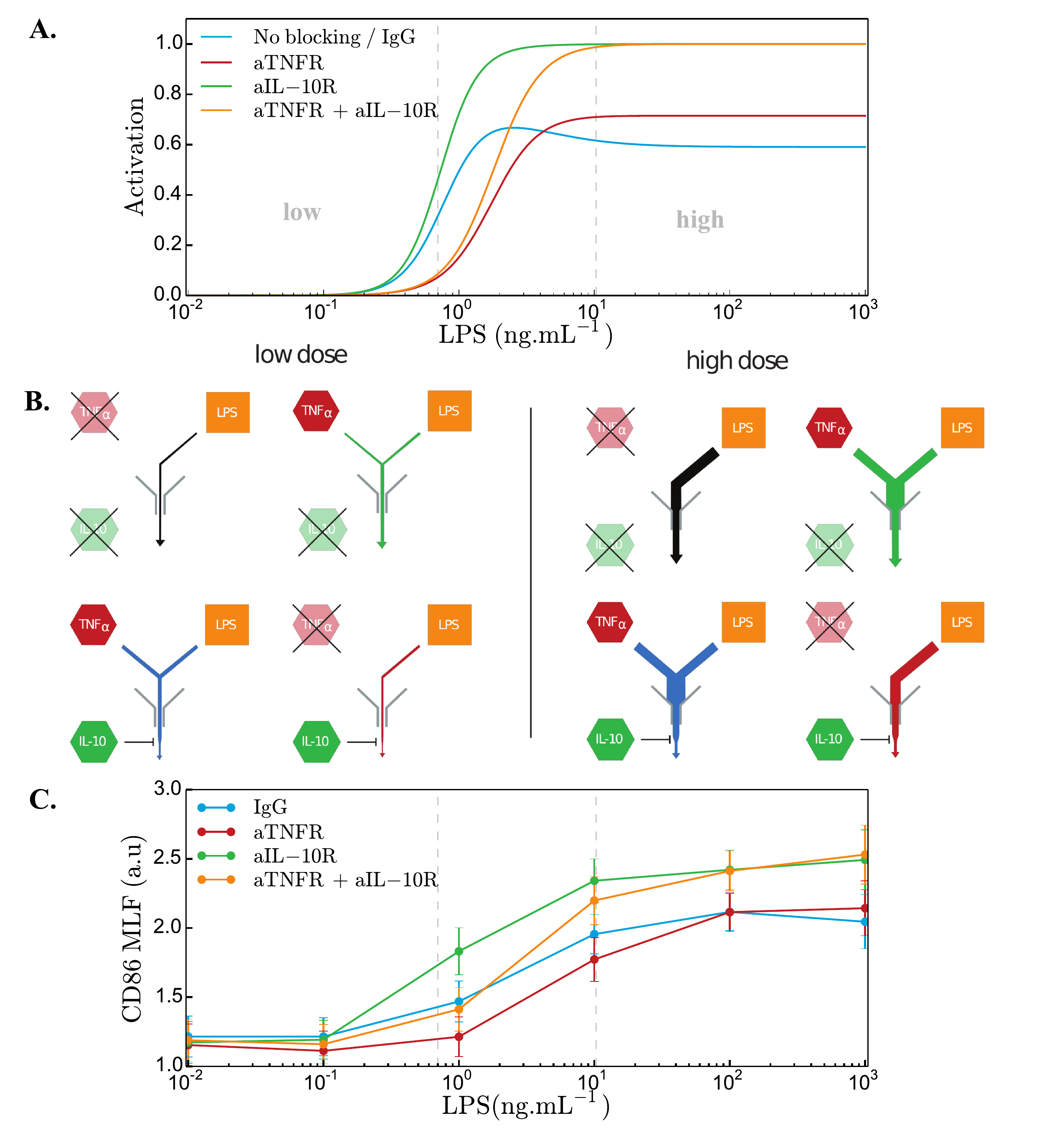}
\caption{
\textbf{Bottleneck model explains DC maturation control by opposing endogenous and exogenous signals.}
\textbf{A.} Fraction of activation of DC population for a range of LPS doses as predicted by the steady state model. Using this model we predict the qualitative behavior of the system when both regulatory loops are not functional.
\textbf{B.} Schematic representation of the outcome of the steady state bottleneck model
\textbf{C.} CD86 mean log-fluorescence (MLF) for a range of LPS stimulation strength. Blocking both regulatory loops grants us a good test of the validity of our model. The model offers good qualitative agreement with the data in every condition.
}
\label{fig4}
\end{figure*}

A schematic representation of the effective regulatory pathway described above is shown in Fig.~\ref{fig3}. A central signal integrator combines the two pro-inflammatory signals, TNF$\alpha$ and LPS, in a single common pathway making this integrator the key regulator of DC decision. The integrator acts as a molecular bottleneck for the pro-inflammatory signals (see Fig.~\ref{fig3}): it responds to increases in the pro-inflammatory signal concentrations only until a certain total concentration. This concentration can be reached either purely by TNF$\alpha$ or purely by LPS, or by their combination (see Fig.~\ref{fig4}B). Above this total concentration, set by the effective $EC_{50}$ parameter, the response is saturated and increasing pro-inflammatory signals has no effect on the input. Without the bottleneck effect of the central integrator, the TNF$\alpha$ and LPS pathways would independently control the CD86 response. In this case blocking the TNF$\alpha$ loop would not change the $EC_{50}$ of the response to LPS, and adding more LPS while the TNF$\alpha$ loop was blocked would lead to an {increased} activation even for infinitely high LPS doses.

The concentration of the integrator molecule controls the amplitude of the response, which is further modulated downstream by the IL-10 anti-inflammatory signal (Fig.~\ref{fig3} and Fig.~\ref{fig4}B). A plumbing analogy helps illustrate the role of the bottleneck and downstream anti-inflammatory regulation: there is a very high source of water distributed to each house, but the amount of available water is limited by the throughput capacity of the main pipeline (this is the bottleneck that regulates the amount of pro-inflammatory signals -- see Fig.~\ref{fig4}B). However when you take a shower, you can regulate the waterflow directly at the faucet (this is the inhibitory action of IL-10). In the absence of IL-10, the bottleneck still limits the scale of the inflammatory response. IL-10 can further downregulate it. 

The bottleneck model reproduces all the experimentally observed features in Fig.~\ref{fig1}B. It further predicts the combined effect of blocking both the IL-10 and TNF$\alpha$ loops (Fig.~\ref{fig4}A). We graphically represent the predictions of the model  for the four blocking conditions at low and high LPS concentrations in Fig.~\ref{fig4}B. At high LPS dose the bottleneck limits the signaling of the master integrator, regardless of whether both TNF$\alpha$ or LPS are sensed or only LPS, and the IL-10 further reduces the strength of the response. At low LPS concentrations the effect of the bottleneck is reduced but IL-10 further modulates the output. We experimentally validated the bottleneck model by blocking both loops simultaneously in LPS stimulated DC. In agreement with the prediction, the condition in which both loops were blocked affected the CD86 IC50 expression similarly to blocking the TNF$\alpha$ loop only in lower LPS dose (Fig.~\ref{fig4}C). At higher LPS doses the CD86 amplitude increased similarly to blocking IL-10 alone, also in agreement with the model predictions. 

\subsection{Paracrine signalling predominantly controls DC maturation}

\begin{figure}[h]
\centering
\noindent\includegraphics[width=\linewidth]{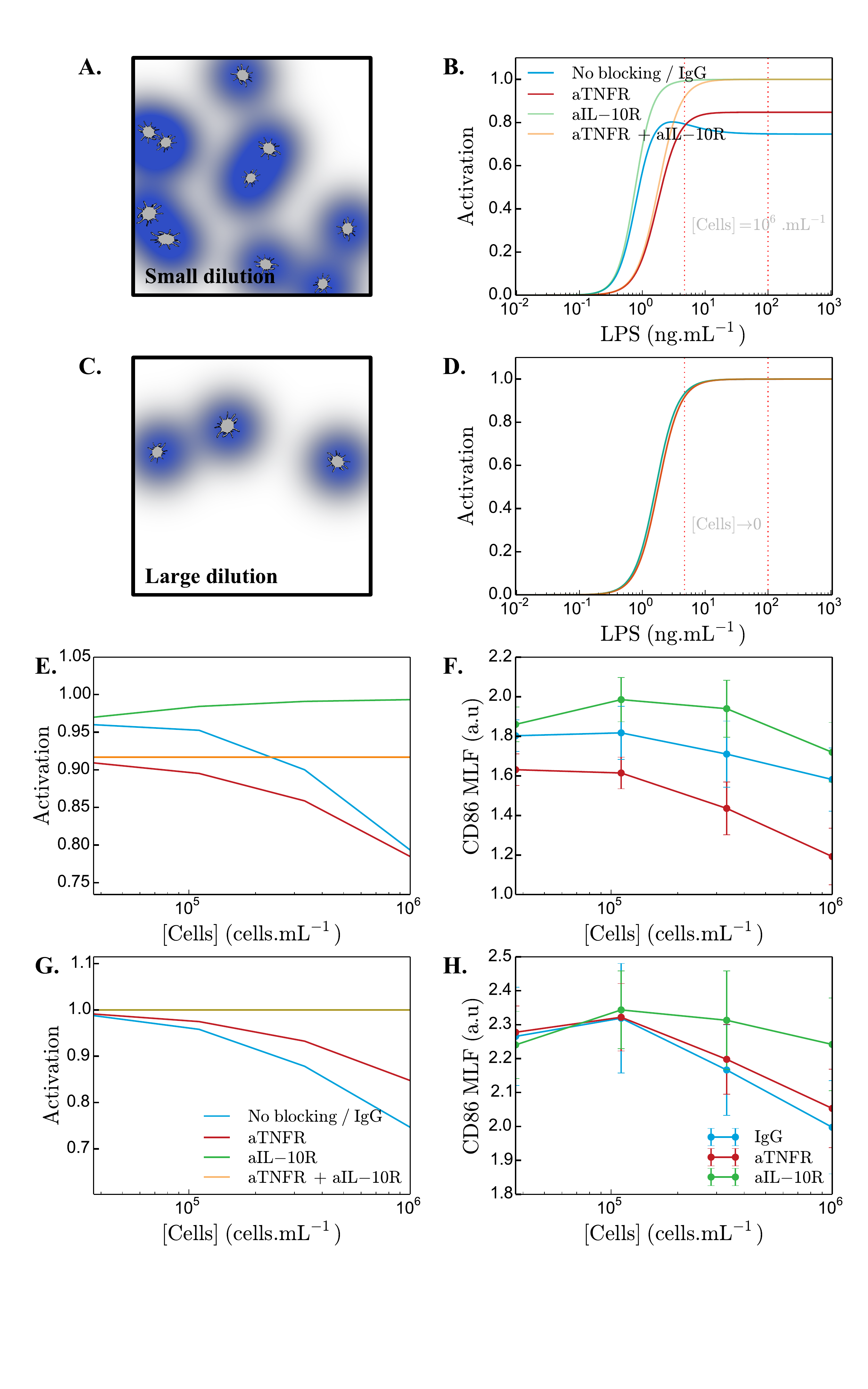}
	\caption{
\textbf{Discriminating autocrine from paracrine loops using dilutions. }
\textbf{A.} and \textbf{B.} Cartoon and prediction of our steady state model with diffusion for high cell concentration.
\textbf{C.} and \textbf{D.}Cartoon and prediction of our steady state model with diffusion for very low cell concentration.
\textbf{E.}Prediction of our steady state model with diffusion on DC activation for a weak LPS stimuli. Computing the expected activation for a range of cell concentrations gives us a qualitative prediction for serial dilutions experiments.
\textbf{F.} Corresponding dilution experiment with low dose of LPS ($\mathrm{1ng.mL^{-1}}$).
\textbf{G.} Model prediction for high dose of LPS.
\textbf{H.} Corresponding dilution experiment with high dose of LPS ($\mathrm{100ng.mL^{-1}}$).
LPS concentration values used for the model predictions in \textbf{E.} and \textbf{G.} are shown with red dashed lines in \textbf{B.} and \textbf{D.}.
	}
\label{fig5}
\end{figure}

\begin{figure}[h]
\noindent\includegraphics[width=\linewidth]{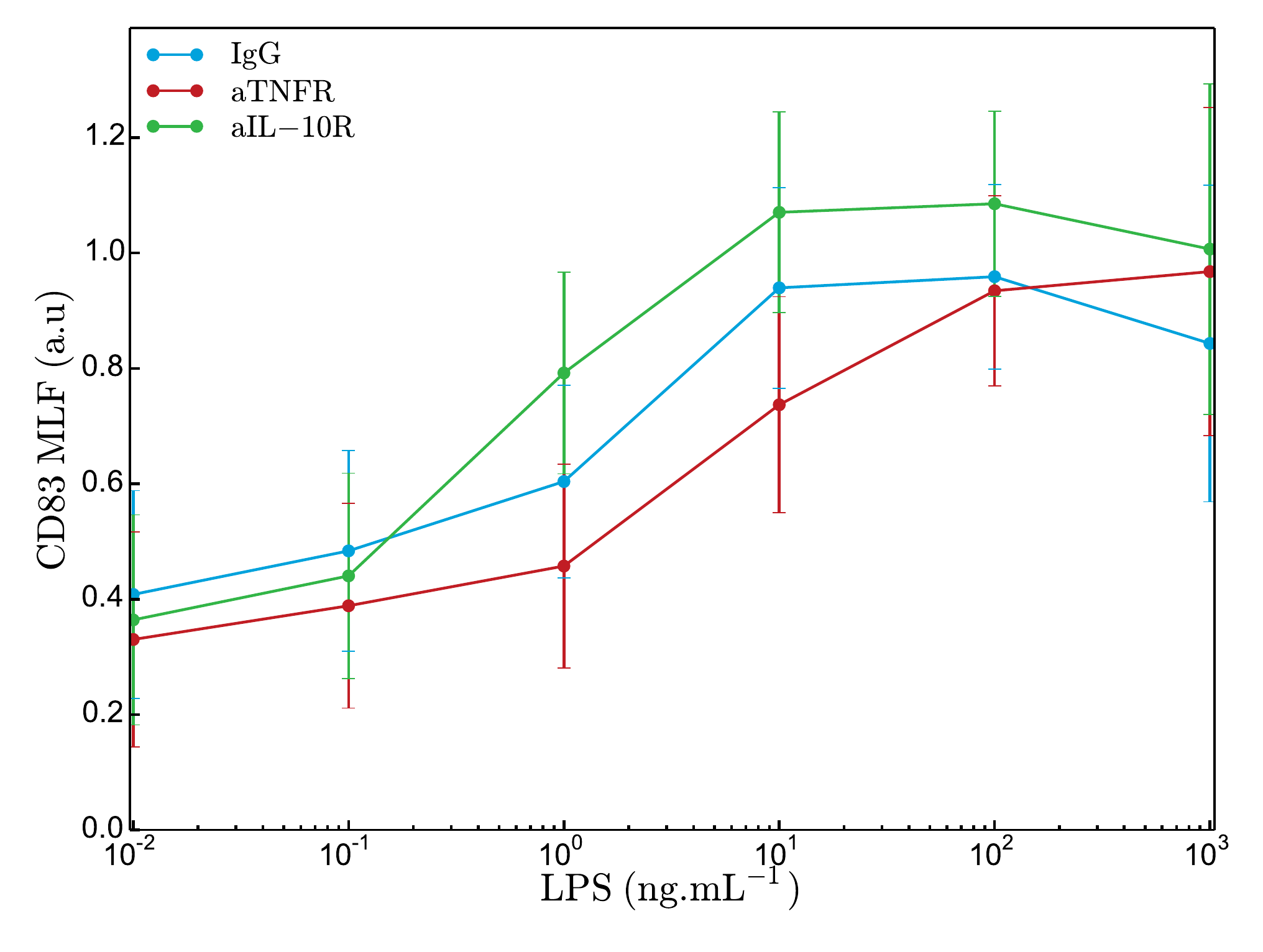}
\caption{
 CD83 mean log-fluorescence (MLF) is shown for a range of LPS concentration incubated with isotypic control (blue), anti-TNFR (red) or anti-IL-10R (green) antibodies.
}
\label{figS1}
\end{figure}

\begin{figure}[h]
\noindent\includegraphics[width=\linewidth]{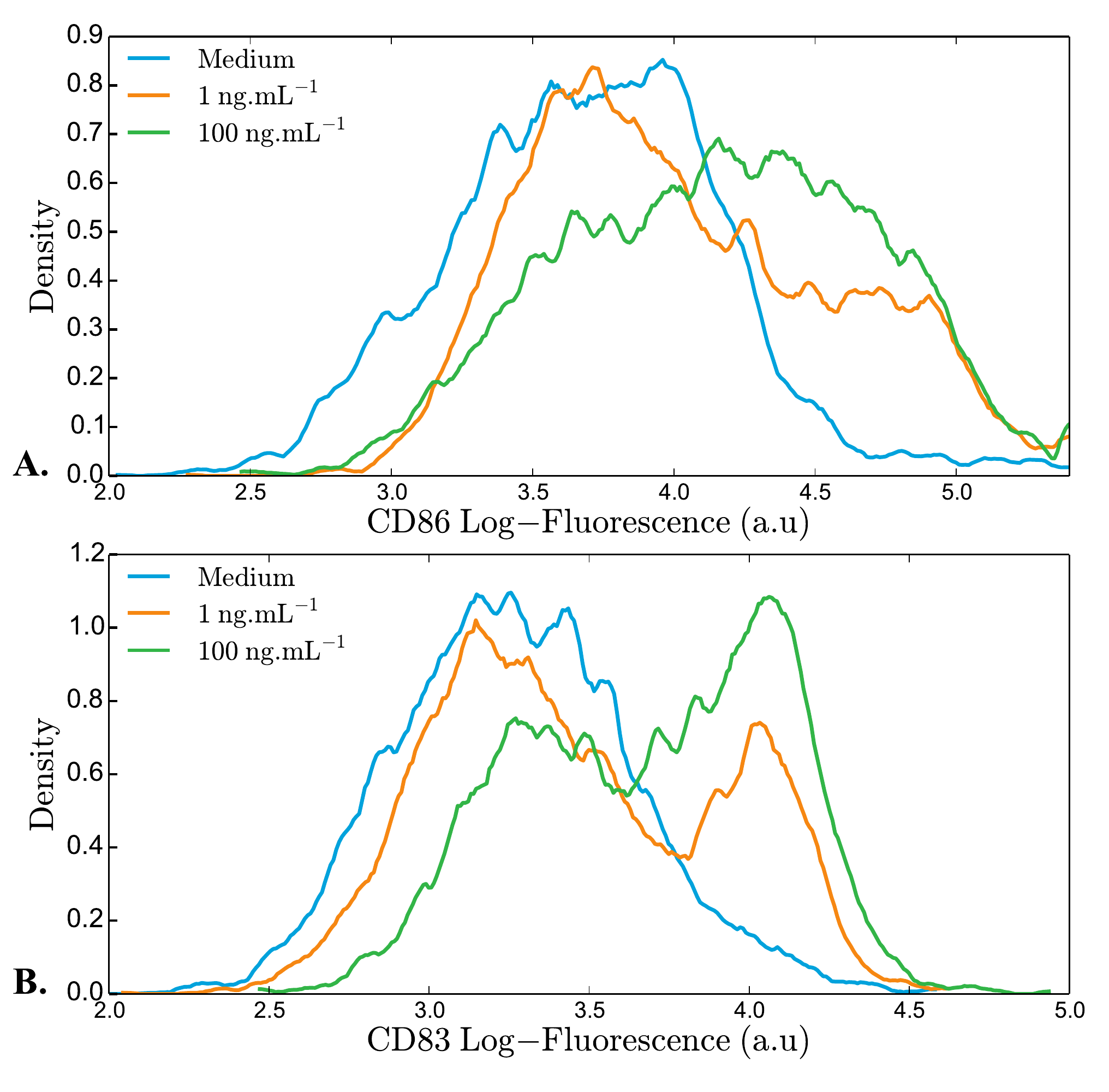}
\caption{
Distribution of CD86 ({\bf A.}) and CD83 ({\bf B.}) log-fluorescence for zero (blue), $1$ ng/ml (orange) and $100$ ng/ml LPS concentrations.
}
\label{figS2}
\end{figure}

DC in our experiment, as in the organism, are not isolated and signal integration depends on the diffusion of cytokines: a cytokine produced by a given cell could be picked up by a receptor on the surface of this same cell (autocrine loop) or by a neighboring cell (paracrine loop). Since we cannot directly measure inter-cellular communication with single molecule resolution, we designed and performed cell dilution experiments to get insight into DC communication at a larger spatial scale. At high cellular concentrations, cells can sense signals from nearby cells (Fig.~\ref{fig5}A), and at large dilutions, only from themselves (Fig.~\ref{fig5}C). Large dilution conditions correspond to pure autocrine signaling. This experiment is based on the assumption that the effect of a purely paracrine loop will decrease as cells are diluted, while a purely autocrine loop will not be sensitive to dilution of the population density. Since the effect of the TNF$\alpha$ feedback loop was observed at low LPS concentrations, whereas the IL-10 feedback was active at high LPS concentrations, we performed dilution experiments at two distinct LPS doses. The role of autocrine and paracrine signaling has recently been addressed in immune cells \cite{douganer2015autocrine}.

To predict the behavior of the DC response in the dilution experiments we combined our phenomenological bottleneck model (Fig.~\ref{fig3}) with diffusion-based estimates for the probabilities of autocrine and paracrine absorption in an effective heterogeneous medium \citep{coppey2007time}(see Materials and Methods for details). Using previously measured kinetic and geometric parameters (see Materials and Methods), the theoretical calculation predicts that a large fraction of the signaling is paracrine in nature. In Fig.~\ref{fig5}E and G we plot the predictions for the mean log CD86 expression at a low and high LPS concentration as a function of the cell concentration. If most of the signaling is paracrine in nature, as we see that at high LPS concentrations (Fig.~\ref{fig5}G), with increasing cell dilutions all the blocking conditions converge to nearly the same activation levels, equal to the levels predicted in the case when all the loops are non-functional (orange curve in Fig.~\ref{fig5}B and G). For very low cell density we expect the paracrine feedback loops to have no effect on CD86 expression and all feedback takes place by autocrine loops. Measurements of ligand affinity of TNF$\alpha$ and IL-10 to their respective membrane receptors \cite{grell1998type,tan1993characterization,liu1994expression} show that TNF$\alpha$ has a greater affinity for its receptor than IL-10 does. We thus predict that TNF$\alpha$ autocrine fraction should be greater than IL-10. Since the effect of the TNF$\alpha$ feedback is observed at low LPS concentrations, and the IL-10 feedback at high LPS concentrations, we expect the convergence of the curves corresponding to different conditions at low LPS concentrations to be less pronounced than at high concentrations. Our model predicts (see Fig.~\ref{fig5}E) that the curves corresponding to blocking the TNF$\alpha$ loop do not converge to those where the TNF$\alpha$ is active at high dilutions for low LPS concentration.

To experimentally assess the effect of dilutions on the loops we activated DC with either low (1 ng/ml) or high (100ng/ml) LPS in different cell dilutions with the initial culture concentration being $\mathrm{10^6{\rm cells.mL}^{-1}}$ (Fig.~\ref{fig5}F and H). In agreement with our model we could observe that at both low and high LPS doses all conditions were converging to the same amount of activation. Because of the saturation effect we could not observe a slower convergence for the case of a blocked TNF$\alpha$ loop for high LPS dose, however it was observable for low LPS dose (Fif\ref{fig5}F). In the case of the lower LPS dose, in which the TNF$\alpha$ loop plays a more specific role, we observed that despite serial dilution, the effect of blocking the loop was maintained, at least to some extent, suggesting the existence of an autocrine signaling. Interestingly, in the higher dose of LPS, the effect of IL-10 loop was rather sensitive to dilutions, suggesting that in a context of high microbial load IL-10 acts in a paracrine manner.

\section{Discussion}

Innate immune recognition is key to promote an efficient anti-microbial immune response, but also needs to be controlled, in order to avoid immunopathology. It is known that immune activating and immune dampening signals are both rapidly produced and co-exist within any inflamed tissue \cite{tamayo2011pro}. However, the interplay between exogenous microbial signals, and endogenous pro- and anti-inflammatory signals has not been formalized in an integrated manner. This is critical to the decision making of the immune response, as it is driven by multiple dynamic signals, conveying different types of information to innate immune cells. By combining experiments with modeling, we showed that the final response of the DC population relies on integrating the initial signal with the induced pro- and anti-inflammatory responses using feedback loops. The integration is based on two steps: first the pro-inflammatory signals are integrated through a bottleneck and then the amplitude of the result is further modulated by the anti-inflammatory signal. The key element of this integration occurs at the signal bottleneck, which controls the effective concentration range ($EC_{50}$) of the response to LPS and limits the maximum pro-inflammatory response. The anti-inflammatory regulation that follows is mostly paracrine, as opposed to the bottleneck integration that has an autocrine component, suggesting that the final response is modulated based on the population level response. 

Bottleneck signal integration in molecular systems have mostly been proposed for the integration of two positive signals. They were suggested as a means for TNF$\alpha$ activation \cite{cheong2011information}. Here we propose that a bottleneck is the essential component in making the decision to the response in the presence of two opposing signals: IL-10 and TNF$\alpha$. Since the negative regulation by IL-10 acts after the bottleneck, it regulates the maximum level of activation, while the positive TNF$\alpha$ acts before the bottleneck thereby affecting the activation threshold. The two opposing signals thus control distinct aspects of the dose response. This feature is independent of the fact that the two signals have opposing effects: the possibility of additional pre- or post-bottleneck regulation would have the same effect on two positive signals.

A natural candidate for this bottleneck integrator is the widely studied \cite{basak2012lessons} nuclear factor NF$\kappa$ B: several studies demonstrated how LPS and TNF$\alpha$ trigger NF$\kappa$ B nuclear translocation \cite{lee2009noisy}. Additionally the saturation effect observed in our data was also seen when looking at NF$\kappa$ B nuclear translocation due to the limited and constant amount of NF$\kappa$ B\cite{Tay2010}.

Additionally to the main modes of signal integration based on the bottleneck and IL-10 repression, TNF$\alpha$ activates IL-10 expression, while IL-10 represses TNF$\alpha$. These secondary interactions do not change the basic flow of signal integration, but are predicted by the model to produce a maximum in the CD86 at intermediate LPS concentrations (Fig.~\ref{fig4}A). Since LPS activates both IL-10 and TNF$\alpha$, repression of TNF$\alpha$ slightly shifts the EC-50 of the response to larger LPS concentrations, while activation of IL-10 results in a larger moderation of the response than in the absence of TNF$\alpha$ for high LPS concentrations.

The presented results are population averages over multiple independent measurements. The fluorescence distributions plotted in {Fig.~S2} show a large heterogeneity in the population, indicating that particular cells can have very different responses. {The error bars indicate the standard error of the mean over multiple experiments}. The measurement noise is impossible to distinguish from the natural heterogeneity of the response in the population. Given this heterogeneity, the mean CD86 response in the double blocked mutant is consistent with the theoretical prediction. 

Dendritic cells often are surrounded by other dendritic cells and, through secreting signaling molecules, communicate with each other to make a decision at the population level. This collective decision making process can help make the right readout in a noisy environment thus reducing response variability as for wound healing \cite{handly2015paracrine}. By sharing their response, cells in a population can confirm initial measurements by sensing the signals that their neighbors secrete. Alternatively, cells could simply use the feedback loops to amplify their own initial signal to accelerate their response. 

Previous experiments have highlighted the difference between population and single cell measurements in TNF$\alpha$ responses \cite{lee2009noisy}. The nature of the signal (paracrine or autocrine) controls the spatial range of the responding cells and determines the lengthscale on which the decision is made. Feedback loops are necessary elements for integrating population-level signals. The signalling range controls whether there is population level averaging, or whether each cell only listens to itself. Here, by using a combination of dilution and fluorescence experiments with modelling, we show that the anti-inflammatory IL-10 signal is paracrine and long range, as opposed to the autocrine and short range pro-inflammatory TNF$\alpha$ signal. Cells rely on local signals to detect bacterial signals, but  integrate anti-inflammatory signals from anywhere in the population to modulate their response. 

Such a localized pro-inflammatory response can be useful in the case of an infection: cells that are further away from the source of the signal do not need to respond. In view of their signalling ranges, autocrine or paracrine feedback loops have different roles: autocrine signaling modifies the strength of response to LPS of the cell itself, while paracrine signaling is used to transmit information to neighboring cells that may not have been exposed to LPS directly. Such a combination of local, excitatory feedback with global, inhibitory regulation has been suggested as a general way to sense differences in spatial concentration profiles, and has been proposed as a mechanism for detecting spatial concentration gradients in the slime mold Dictyostelium \cite{jilkine2011comparison,iglesias2002modeling,levchenko2002models}, or more recently {for wound healing~\cite{handly2015paracrine} and} in the context of morphogenesis of mammary epithelial cells in response to a gradient of the Epidermal Growth Factor \cite{ellison2016cell}. Our results extend this concept to the immune system following innate microbial sensing.

In summary, in this joint experimental and theoretical study we quantified how a cell makes decisions about the appropriate response to a given concentration of the bacterial signal LPS in the environment, and as a result whether to initiate an inflammatory response or not. More broadly, the mechanisms described give a way to integrate information and make decisions in the presence of conflicting signals.
{Furthermore we show how simple biophysical models give us insights into cell-cell communication in cell density regimes that are inaccessible by single-cell microscopy~\cite{Youk2014, Youk2014a}.}

\section{Materials and Methods}

\subsection{Monocyte-derived dendritic cells generation and activation}

Fresh blood samples collected from healthy donors were obtained from H\^{o}pital Crozatier \'{E}tablissement Fran\c{c}ais du Sang (EFS), Paris, France, in conformity with Institut Curie ethical guidelines. PBMCs were isolated by centrifugation on a Ficoll gradient (Ficoll-Paque PLUS, GE Healthcare Life Sciences). Monocytes were selected using antibody-coated magnetic beads and magnetic columns according to manufacturer’s instructions (CD14 MicroBeads, Miltenyi Biotec). To generate immature DC, CD14+ cells were cultured for 5 days with IL-4 (50 ng/mL) and GM-CSF (10 ng/mL) in RPMI 1640 Medium, GlutaMAX (Life Technologies) with 10\% FCS. Monocyte-derived DC were pre- treated for one hour with mouse IgG1 (20 µg/mL, R\&D Systems), mouse anti-IL10R blocking antibody (10 $\mu$ g/mL, R\&D Systems) or mouse anti-TNF$\alpha$ Receptors 1 and 2 (10 $\mu$ g/mL, R\&D Systems) and then cultured with medium or LPS ( Invivogen) for additional 24 hours. 
Cells were stained for 15 min at 4$\degree$ C using a PE-anti-Human-CD86 (clone 2331, BD), or with the corresponding isotype. Cells were analyzed on a Fortessa instrument (BD Biosciences).

\subsection{Dilution experiments}

The dilution experiments were carried out following the same procedure as described above, but with {1:3, 1:9 and 1:27 fold dilutions}.

\subsection{Phenomenological bottleneck model}

The bottleneck model for signal integration presented in Fig.~\ref{fig3} and described in the main text predicts the concentration $C$ of the differentiation marker CD86 in terms of the concentration $L$ of LPS, and the concentration $I$  of IL-10 and $T$ of TNF$\alpha$, expressed in fraction of their maximal values. Note that IL-10 and TNF$\alpha$ concentrations depend on $L$ themselves.
The experimental observations are summarized by the minimal phenomenological model:
\begin{align} 
T & =\frac{1}{1+I/(1-\beta)}\times\frac{1}{1+({K_{T}}/{L})^{n_{T}}},\label{eq1}\\
I & =\frac{1-\alpha+\alpha T}{1+({K_{I}}/{L})^{n_{I}}},\label{eq2}\\
B & =\frac{1}{1+\left(\frac{K_{C}}{L+\theta {T}}\right)^{n_{C}}},\label{eq3}\\
C & =\frac{B}{1+I},\label{eq4}
\end{align}
where $K_{T}$, $K_{I}$ describe the concentrations of LPS at which TNF$\alpha$ and IL-10, respectively, would reach half maximal activation in the absence of other regulation. $n_{T}$ and $n_{I}$ are Hill coefficients describing the effective steepness of the regulation function. Note that, in addition to being regulated by LPS, IL-10 and TNF-$\alpha$ also regulate each other (Eq.~\ref{eq1} and \ref{eq2}). The strength of this mutual interaction is set by the constants $\alpha$ and $\beta$.
The pro-inflammatory signals from LPS and LPS-activated TNF$\alpha$ are then integrated by activation of a common intermediate species of concentration $B$ (Eq.~\ref{eq3}), which reaches half-maximal expression at a joined input concentration of $K_{C}$ with an effective cooperativity of $n_{C}$. $\theta$ describes the maximal concentration of TNF$\alpha$, so that TNF concentration is given by $\theta T$. $B$ is given in arbitrary units so that its maximum activation is $1$ at saturation.
Signal integration through $B$ creates a signaling bottleneck:  at low levels of pro-inflammatory signals the response $B$ is proportional to the sum of both LPS and TNF$\alpha$ signals, $L$ and $\theta T$, but at high levels large concentrations of either LPS or TNF$\alpha$ are sufficient to saturate the response. The final concentration of CD86, $C$ (Eq.~\ref{eq4}) is modulated downstream by the anti-inflammatory IL-10 that represses the integrated pro-inflammatory signal, modulating the amplitude of the response but not the concentration range of the response. The model is a phenomenological model that captures the relevant features of the complex interaction network by focusing on the effective interactions. The parameters, such as the concentrations at half maximal expression, do not describe individual mechanistic molecular interactions, but the effective outcome of many such reactions.

To calculate the output concentration $C$ as a function of $L$, $I$ and $T$ are first determined self-consistently by solving the system of two equations \ref{eq1} and \ref{eq2} for a given $L$. Then their values are injected into Eq.~\ref{eq3} and \ref{eq4} to obtain $C$.

\subsection{Ligand diffusion model}

Each secreted signaling molecule, TNF$\alpha$ or IL-10, can have four distinct fates. He can be absorbed by the cell that emitted it with probability $P_{\rm auto}$ (autocrine signaling), or by another cell with probability $P_{\rm para}$ (paracrine signaling). It can be degraded in the medium with probability $P_{\rm deg}$, or left free in solution with probability $P_{\rm free}$.

\begin{figure}[h]
\noindent\includegraphics[width=\linewidth]{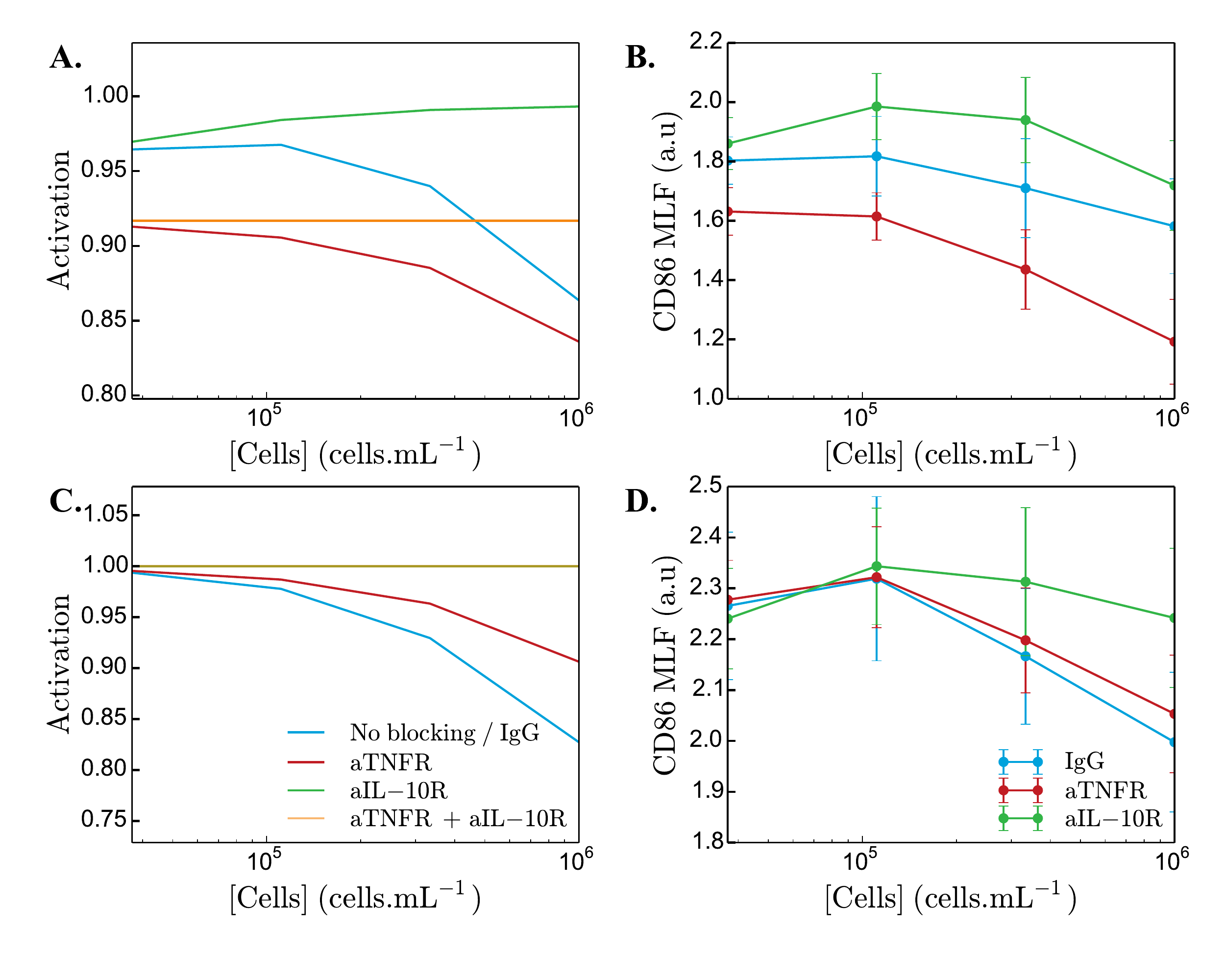}
\caption{
Diffusion model prediction for 5h of effective measurement.
}
\label{figS3}
\end{figure}

{To estimate the role of autocrine and paracrine signalling, we use the results derived by Coppey et al \cite{coppey2007time} for the probability of autocrine absorption of a ligand.
This result assumes an effective rate with which the secreted molecule encounters another cell and is absorbed by it, ${k_{\rm abs}}={4\pi D R^{2}\kappa}/{(D+R\kappa}c)$. This rate depends on the cell concentration $c$, the ligand diffusivity $D$, and on an effective surface trapping rate $\kappa={{k_{\rm on}}N_{R}}/{4\pi R^{2}}$, where $R$ is the cell radius, $N_{R}$ the number of receptors displayed at the surface of the cell and ${k_{\rm on}}$ the rate constant of the ligand-receptor binding.
In addition, molecules are degraded with rate $k_{\rm deg}$, resulting in an effective rate $k_b=k_{\rm abs}+k_{\rm deg}$ of molecules disappearing from the medium.

At long times, the probability of autocrine signaling then reads:
\begin{equation}\label{eq_pauto}
P_{\rm auto} =\frac{{\kappa}}{\frac{D}{R}+{\kappa}+\sqrt{D k_b}}.
\end{equation}

This result is valid for times that are longer than the typical time it takes for the ligand to be recaptured by the cell that secreted it. In that regime, which is that of the measurements, 
\begin{align}
P_{\rm free}(t)&=e^{-k_b t}(1-P_{\rm auto}),\\
P_{\rm para}(t)&=\frac{k_{\rm abs}}{k_b}(1-e^{-k_b t})(1-P_{\rm auto}),\\
P_{\rm deg}(t)&=\frac{k_{\rm deg}}{k_b}(1-e^{-k_b t})(1-P_{\rm auto}),
\end{align}
where we have used $P_{\rm abs}/P_{\rm deg}=k_{\rm abs}/k_{\rm deg}$, $P_{\rm auto}+P_{\rm para}+P_{\rm free}+P_{\rm deg}=1$, and the fact that ligands having escaped autocrine absorption have probability $e^{-k_b t}$ of staying in the medium (i.e. not being absorbed by another cell or degraded) after time $t$.

The fraction $f$ of ligands that contribute to signaling after time $t$ is thus the sum of the probabilities of autocrine and paracrine signaling,
\begin{equation}\label{f(c)eq}
f = {P_{\rm auto}} + {P_{\rm para}}.
\end{equation}
This fraction depends on both the time $t$ since the beginning of the experiment, and the concentration $c$ of cells. Since IL-10 and TNF$\alpha$ signaling may have different properties, we define two fractions $f_I$ and $f_T$ corresponding to the fraction of IL-10 and TNF$\alpha$ contributing to signaling.

Equations \eqref{eq1}-\eqref{eq4} are then modified to:
\begin{align} 
T & =\frac{1}{1+f_II/(1-\beta)}\times\frac{1}{1+({K_{T}}/{L})^{n_{T}}},\label{eqd1}\\
I & =\frac{1-\alpha+\alpha f_T T}{1+({K_{I}}/{L})^{n_{I}}},\label{eqd2}\\
C & =\frac{1}{1+\left(\frac{K_{C}}{L+\theta f_T {T}}\right)^{n_{C}}}\times \frac{1}{1+f_II}.\label{eqd3}
\end{align}
Each quantity depends on the cell concentration $c$ and on the time $t$ through $f_I$ and $f_T$.

{In order to probe the behavior of the system we gathered plausible model parameters from the literature. For a number of parameters precise estimates were not found and we used analogue estimates for other systems (e.g macrophages). These parameters are presented and justified in the following section and summarized in Table S1. Because of our focus on the phenomenology of the model, we did not try to make precise numerical predictions, but rather give qualitative ones. Parameters were not fitted to the data. They were either taken from the literature, or picked to reproduce the data qualitatively.}
The rate constant for TNF$\alpha$/TNFr1 and TNF$\alpha$/TNFr2 association is estimated to be $10^7-10^8{\rm M}^{-1}{\rm s}^{-1}$ \cite{grell1998type} and the IL-10 association rate to $10^5-10^6 {\rm M}^{-1}{\rm s}^{-1}$ \cite{tan1993characterization,liu1994expression}. We took a cell radius of $25 \mu{\rm m}$, an initial cell concentration of $10^6 {\rm mL}^{-1}$ before dilution, a diffusion coefficient $D = 10^2\mu {\rm m}^2{\rm s}^{-1}$
\cite{coppey2007time}, and the number of receptors $N_R= 10^3-10^4$ (the number of receptors on dendritic cells is not known and we use an order of magnitude estimate based on macrophages \cite{liu1994expression,imamura1987expression,pocsik1994cell,weber1994lipopolysaccharide,ding2001differential}).
While TNF degradation rate was measured to be $\mathrm{2 \times 10^{-4} {\rm s}^{-1}}$ \citep{Tay2010}, interleukins are classically described as extremely stable, meaning that measured degradation rates actually correspond to absorption by other cells.
We thus assumed $k_{\rm deg}^{IL-10}$ to be negligible.
When DCs were cultured for 24 hours, the delay with which DCs start signalling and the fact that DCs make decision on a shorter timescale should be considered. We picked the actual signal integration time $t$ to be lower than 24h and to be around 10h. Changing this integration time to $t=5$h (Fig.~\ref{figS3}) did not affect the model prediction.

{These parameter values allow us to estimate that the majority of the effective signaling is paracrine. TNF$\alpha$α has a stronger autocrine component, $P_{\rm auto}^{\rm{\text{TNF}}}/(P_{\rm auto}^{\rm{\text{TNF}}}+P_{\rm para}^{\rm{\text{TNF}}})=5\text{--}18\%$, than IL-10, $P_{\rm auto}^{\rm{\text{IL-10}}}/(P_{\rm auto}^{\rm{\text{IL-10}}}+P_{\rm para}^{\rm{\text{IL-10}}})={0.1\text{--}2.3\%}$, in the range of concentration experimentally tested}.

We used the estimates of the probability of auto- vs paracrine signaling combined with our bottleneck model to predict the response of CD86 at different dilutions. We simulated a population of cells in a medium with low ($L=4.7{\rm ng}\cdot {\rm L}^{-1}$) and high ($L=100{\rm ng}\cdot {\rm L}^{-1}$) LPS concentrations and used the properties of autocrine and paracrine signaling derived above to estimate the population level response at different dilution. 
The parameters used for the diffusion model and the corresponding references are listed in Table~S1.}

\section{Acknowledgements}

This work was supported by  MCCIG grant no. 303561, by the Agence Nationale pour la Recherche (ANR), by the Fondation pour la Recherche Medicale (FRM), by the European Research Council (ERC Starting Grant 306312 and ERC Consolidator Grant 281987), by ANR-10-IDEX-0001-02 PSL, ANR-11-LABX-0043, CIC IGR-Curie 1428, and by EMBO and Institut Curie post-doctoral fellowships to ICL.

\bibliographystyle{pnas}

\newpage

\beginsupplement
\section{Supplementary Information}

\begin{table}[h]\label{table_parms}
\centering
\begin{tabular}{|c|c|c|}
	\hline
	Parameter & Value & Ref\\
	\hline
	\hline
	$\beta$ & 0.6 & -\\
	$\alpha$ & 0.6 & -\\
	$\vartheta$ & 20& -\\
	$K_{T}$ & 5 & -\\
	$K_{I}$ & 0.8 & -\\
	$K_{C}$ & 1.8 & -\\
	$n_{T}$ & 1.5 & -\\
	$n_{I}$ & 1.5 & -\\
	$n_{C}$ & 2.5 & -\\
	$D$ & $\mathrm{10^{-10}\ m^2.s^{-1}}$ & \cite{coppey2007time}\\
	$R$ & $\mathrm{25\ \mu m}$ &  \cite{coppey2007time}\\
	$N_R$ & $\mathrm{10^4}$ & \cite{liu1994expression,imamura1987expression,pocsik1994cell,weber1994lipopolysaccharide,ding2001differential}\\
	$k_{\rm on}^{TNF\alpha}$ & $\mathrm{10^{8}\ M^{-1}.s{-1}}$ & \cite{grell1998type}\\
	$k_{\rm on}^{IL-10}$ & $\mathrm{10^{6}\ M^{-1}.s{-1}}$ & \cite{tan1993characterization,liu1994expression}\\
	$k_{\rm deg}^TNF$ & $\mathrm{2 \times 10^{-4}.s^{-1}}$ & \cite{Tay2010}\\
	$k_{\rm deg}^{IL-10}$ & 0 & -\\
	\hline	

\end{tabular}
\caption{Model parameters table}
\end{table}

\end{document}